\documentclass[12pt]{article}
\usepackage{graphicx}
\usepackage{dcolumn}
\usepackage{bm}
\usepackage{amsmath}
\usepackage{amssymb}
\usepackage{color}
\usepackage{hyperref} 
\title{Quantum statistical mechanics from a Bohmian perspective}
\author{
Hrvoje Nikoli\'c \\
Theoretical Physics Division, Rudjer Bo\v{s}kovi\'{c} Institute, \\
P.O.B. 180, HR-10002 Zagreb, Croatia \\
{\normalsize e-mail: hnikolic@irb.hr} 
}
\date{\today}
\begin{document}
\maketitle
\begin{abstract}
We develop a general formulation of quantum statistical mechanics
in terms of probability currents that satisfy continuity equations
in the multi-particle position space, for closed and open systems with a fixed number of particles.
The continuity equation for any closed or open system suggests a natural Bohmian interpretation
in terms of microscopic particle trajectories, that make the same measurable predictions
as standard quantum theory. 
The microscopic trajectories are not directly observable,
but provide a general, simple and intuitive microscopic interpretation of macroscopic phenomena 
in quantum statistical mechanics. In particular, we discuss how various notions of entropy, 
proper and improper mixtures, and thermodynamics are understood from the Bohmian perspective.  
\end{abstract}

\vspace*{0.5cm}

\section{Introduction} 

Bohmian mechanics \cite{bohm,book-bohm,book-hol,book-durr,oriols} is an interpretation of quantum mechanics (QM)
in terms of variables that always have definite values, irrespective of whether they are measured or not.
In the simplest and best known form, these variables are positions of pointlike particles, which 
change with time in a deterministic manner. According to this interpretation, all quantum uncertainties and probabilistic laws 
emerge similarly as in classical statistical physics, from a practical lack of knowledge of the actual 
initial positions. Very general analysis \cite{bohm,book-bohm,book-hol,book-durr,oriols} shows that Bohmian mechanics 
makes the same measurable predictions as standard QM. Thus, in principle, 
all quantum phenomena that can be described by standard QM, can also be explained by Bohmian mechanics.

Since Bohmian interpretation 
offers an explanation of the origin of probabilities, it seems natural to study, 
in more detail, how Bohmian mechanics explains the phenomena that otherwise are described by standard
quantum statistical mechanics \cite{huang,reif,kardar1}.
Surprisingly, however, very little is done in that direction. Bohm and Hiley \cite{book-bohm,bohm-hiley}
studied quantum statistical mechanics in the ``Bohmian'' context by studying only the statistics of wave functions,
without saying much on the role of actual particle positions in understanding statistical mechanics.
Goldstein and his coworkers, who are leading experts in both Bohmian mechanics and statistical mechanics, 
published very little on the role of Bohmian mechanics in quantum statistical mechanics. 
In 1992 \cite{goldstein_absolute_uncertainty} they concluded that deriving quantum statistical mechanics 
from Bohmian mechanics ``has not yet reached its infancy".
In 2005 \cite{goldstein-density-matrix} they studied how Bohmian trajectories can be defined for mixed density operators.
In 2020 \cite{goldstein-BoltzGibbs-CQ} they studied thoroughly Gibbs and Boltzmann entropy in classical and quantum mechanics,
by saying very little on the role of Bohmian trajectories in that context.
Bricmont, who is also an expert in both quantum foundations \cite{bricmont-sense,bricmont-sense-nonsense}
and statistical mechanics \cite{bricmont-statist},
and is a strong supporter of the Bohmian interpretation \cite{bricmont-bohm},
in his fantastic monograph on foundations of statistical mechanics \cite{bricmont-statist}
studied only classical statistical mechanics.
 
It is not clear to us why, so far, foundations of quantum statistical mechanics have not been studied more thoroughly
from a Bohmian perspective. 
Be that as it may, in this paper we 
try to fill the gap by presenting a systematic investigation of foundations
of quantum statistical mechanics from the Bohmian point of view. 
To our knowledge, this is the first (hopefully not the last) work aimed specifically at this topic.
A large part of our work, however, contains novel results in foundations of statistical mechanics
that do not depend on the Bohmian interpretation, which can also be of value in foundations of 
statistical mechanics from the point of view of standard QM.
Indeed, our view is not that the Bohmian interpretation is ``the right'' interpretation that should 
replace standard QM. Instead, our view is that Bohmian mechanics 
is an auxiliary practical thinking tool \cite{nikIBM} that {\em enriches} standard QM.   

We start with foundations of classical statistical mechanics in Sec.~\ref{SECcsm}, 
in closed and open systems, 
with emphasis on probability densities and probability currents
satisfying continuity equations in multi-particle phase spaces.
In particular, we develop a general systematic method of averaging 
over the environment that defines a continuity equation for any open 
subsystem with a fixed number of particles. 

In analogy with classical probability currents in phase space, 
in Sec.~\ref{SECqstat} we develop in detail a similar
theory of quantum probability densities and probability currents, for closed and open systems,  
that satisfy continuity equations in multi-particle position spaces.
The averaging over the environment, as well as ``hiding" the spin degrees of freedom
within the currents in position space,
are treated on an equal footing through an elegant formalism of partial tracing.

The Bohmian interpretation is introduced in Sec.~\ref{SECbohm}, where,
in analogy with classical particle velocities in multi-particle phase space, 
the quantum particle velocities are defined by the quantum probability currents
in multi-particle position space. Even though the particle currents 
for closed system and open subsystem define different particle trajectories, 
both theories make the same measurable predictions in the open subsystem.
Hence, for all practical purposes, one can formulate a Bohmian 
interpretation of open subsystem that does not depend on particle positions
in the environment. In this interpretation, the formula for Bohmian velocities
in the open subsystem
is equivalent to the formula in \cite{goldstein-density-matrix} 
for Bohmian velocities for the ``fundamental density matrix", except that 
in our case the density matrix is not fundamental (because it only describes 
a subsystem), and we use a different (hopefully more physicist friendly) notation.
  
The results and insights above are applied in Sec.~\ref{SECexplssm}
to better understand the general principles of quantum statistical mechanics 
from a Bohmian point of view. We explain how thermodynamics 
is explained by Bohmian mechanics, discuss the difference between proper 
and improper mixtures, and emphasize that quantities such as energy and entropy 
in quantum statistical mechanics are at least partially nomological (i.e., law like), 
rather than purely ontological. 
In particular, we study in detail the conceptual 
differences between different kinds of entropy, namely von Neumann entropy, quantum 
Boltzmann entropy, as well as their classical cousins Gibbs entropy and classical 
Boltzmann entropy. In the analysis of entropies we are particularly influenced 
by the work of Goldstein and his collaborators 
\cite{goldstein-BoltzGibbs-C,goldstein-BoltzGibbs-CQ}
(for other discussions of the relation between Gibbs and Boltzmann entropy 
see also \cite{jaynes,kuic,bain}), but our conclusions are
slightly different because we do not think that there is such thing as ``true" entropy. 
In the special case of thermal equilibrium,
the statistical entropy is required to correspond to thermodynamic entropy,
but this requirement alone is not sufficient 
to uniquely define entropy out of equilibrium, so there is a lot of freedom to
define statistical entropy in various ways which are not equivalent out of equilibrium.   

The conclusions are drawn in Sec.~\ref{SECconc}.

\section{Classical statistical mechanics}
\label{SECcsm}

\subsection{Classical statistical mechanics of a closed system}
\label{SECclasclos}

In this subsection we present a review of general foundations of classical statistical mechanics of closed systems
(for more details see, e.g., \cite{huang,reif,kardar1}) that are needed for a thorough understanding of the rest of the paper.

Consider a closed system of $N$ classical particles with positions ${\bf x}_a$ and momenta  ${\bf p}_a$, 
$a=1,\ldots, N$. It is convenient to introduce a compact notation $x=({\bf x}_1,\ldots,{\bf x}_N)$,
$p=({\bf p}_1,\ldots,{\bf p}_N)$, $z_a=({\bf x}_a,{\bf p}_a)$, $z=(z_1,\ldots,z_N)$, 
where $z_a$ is a point in the 6-dimensional phase space of the $a$'th particle, and $z$ is a point in 
the full $6N$-dimensional phase space of all the particles. The particles have trajectories 
${\bf x}_a(t)$, ${\bf p}_a(t)$ given by the Hamilton equations of motion
\begin{equation}
 \frac{d{\bf x}_a}{dt}=\frac{\partial H}{\partial {\bf p}_a} , \;\;\;
\frac{d{\bf p}_a}{dt}=-\frac{\partial H}{\partial {\bf x}_a} ,
\end{equation}
where $H(x,p)=H(z)$ is the Hamiltonian of the closed system.
This motivates us to define the velocity ``field'' $v_a(z)$ in phase space
\begin{equation}\label{va}
 v_a \equiv \left( \frac{\partial H}{\partial {\bf p}_a} , -\frac{\partial H}{\partial {\bf x}_a} \right) ,
\end{equation}
so that the Hamilton equations of motion can be written more compactly as
\begin{equation}\label{eomclas}
 \frac{dz_a}{dt} = v_a.
\end{equation}

Now consider the probability density $\rho(z,t)=\rho(x,p,t)$ in the full phase space.
Clearly we have
\begin{eqnarray}\label{preLiouville}
 \frac{d\rho}{dt} &=& \frac{\partial\rho}{\partial t} 
+ \sum_a \left( \frac{\partial\rho}{\partial {\bf x}_a} \frac{d{\bf x}_a}{dt} 
+ \frac{\partial\rho}{\partial {\bf p}_a} \frac{d{\bf p}_a}{dt}     \right) 
\nonumber \\
&=& \frac{\partial\rho}{\partial t} 
+ \sum_a \left( \frac{\partial\rho}{\partial {\bf x}_a} \frac{\partial H}{\partial {\bf p}_a} 
- \frac{\partial\rho}{\partial {\bf p}_a}   \frac{\partial H}{\partial {\bf x}_a}   \right) 
\nonumber \\
&=& \frac{\partial\rho}{\partial t} + \{\rho,H\}_{\rm PB} ,
\end{eqnarray}
where $\{\;,\;\}_{\rm PB}$ denotes the Poisson bracket. 
Since the particles in a statistical ensemble\footnote{The number of systems in the statistical ensemble
(namely, the number of elements in the set of system's replicas identically prepared in a statistical sense)
is usually taken to be infinite.
This number has nothing to do with the number of particles $N$ in the system, which can be either large or small.}
just follow their trajectories,
the probability density satisfies the continuity equation in the phase space
\begin{equation}\label{contclas}
 \frac{\partial\rho}{\partial t}+\sum_a \nabla_a j_a =0 ,
\end{equation}
where 
\begin{equation}\label{current}
j_a=\rho v_a 
\end{equation}
is the {\em probability current}.
Since 
\begin{equation}
 \sum_a \nabla_a j_a = \sum_a (\nabla_a\rho) v_a + \rho \sum_a \nabla_av_a ,
\end{equation}
and since
\begin{eqnarray}\label{nablav}
 \sum_a \nabla_av_a &=& \sum_a \left( \frac{\partial}{\partial{\bf x}_a} \frac{d{\bf x}_a}{dt} 
+ \frac{\partial}{\partial{\bf p}_a} \frac{d{\bf p}_a}{dt} \right)
\nonumber \\
&=& \sum_a \left( \frac{\partial}{\partial{\bf x}_a} \frac{\partial H}{\partial {\bf p}_a}
- \frac{\partial}{\partial{\bf p}_a} \frac{\partial H}{\partial {\bf x}_a} \right) =0,
\end{eqnarray}
it follows that 
\begin{equation}
 \sum_a \nabla_a j_a = \sum_a (\nabla_a\rho) v_a 
=  \sum_a \left( \frac{\partial\rho}{\partial {\bf x}_a} \frac{d{\bf x}_a}{dt} 
+ \frac{\partial\rho}{\partial {\bf p}_a} \frac{d{\bf p}_a}{dt} \right) . 
\end{equation}
Inserting this into (\ref{contclas}) and comparing with (\ref{preLiouville}), 
we see that 
\begin{equation}\label{dot=0}
  \frac{d\rho}{dt}=0.
\end{equation}
Thus (\ref{preLiouville}) reduces to
\begin{equation}\label{Liouville}  
\frac{\partial\rho}{\partial t} = \{H,\rho\}_{\rm PB}, 
\end{equation}
which is called the Liouville equation.

The results above are generally referred to as Liouville theorem. More specifically, by Liouville theorem,
people sometimes mean the result (\ref{nablav}), sometimes the result (\ref{dot=0}), and sometimes
the result that the phase-space volume is conserved, which can be derived either from 
(\ref{dot=0}), or from the ``fluid incompressibility'' (\ref{nablav}). 
However, the continuity equation (\ref{contclas}) by itself does not imply the conservation of 
phase-space volume, so (\ref{contclas}) should not be thought of as a version of ``Liouville theorem''.
Nevertheless, the continuity equation (\ref{contclas}) is sufficient for a consistent 
statistical interpretation; the validity of the Liouville theorem is not necessary in order to have a  
consistent statistical interpretation of the probability density $\rho(z,t)$. 
In general, in classical statistical physics, the average value of the 
function
%
$O(z)$ is given by 
\begin{equation}\label{O}
\bar{O}(t) = \int dz\, O(z)\rho(z,t) , 
\end{equation}
and consistency of it does not depend on validity of the Liouville theorem. 
The Liouville theorem can be used for an additional justification for working in the phase space 
(rather than, for instance, in the position space), but it seems that a fully 
convincing and generally accepted version of that justification does not exist \cite{bricmont-statist}.

Finally, since the continuity equation will turn out to play a central role in this paper,
it is convenient the write the continuity equation (\ref{contclas}) in a compact form
\begin{equation}\label{contclas2}
 \frac{\partial\rho}{\partial t}+\nabla j =0 . 
\end{equation}

\subsection{Classical statistical mechanics of a subsystem}
\label{SECclassub}

Let us split the full phase space into two subspaces, called $A$ and $B$. They contain 
$N_A$ and $N_B$ particles, respectively, such that $N_A+N_B=N$.
The numbers of particles and the Hamiltonian (i.e., the interactions between the particles) can be arbitrary, 
the analysis in this subsection will not depend on this. 
Due to interactions, any subsystem $A$ or $B$ is an open system, so this subsection is really a theory of open systems.
Thus we write $\rho(z,t)=\rho(z_A,z_B,t)$ and the
continuity equation (\ref{contclas2}) can be written as 
\begin{equation}\label{contclasAB}
 \frac{\partial\rho}{\partial t}+\nabla_A j_A + \nabla_B j_B =0 ,
\end{equation}
with a self-explaining notation. 

Now comes a crucial step, which contains the central idea of this subsection, and indeed, of this whole paper.
The idea is to integrate (\ref{contclasAB}) over $dz_B$
\begin{equation}\label{contclasA}
 \int dz_B \frac{\partial\rho}{\partial t}+ \int dz_B \nabla_A j_A + \int dz_B \nabla_B j_B =0 ,
\end{equation}
where of course 
\begin{equation}
dz_B=dx_Bdp_B=d^3x_1\cdots d^3x_{N_B} d^3p_1\cdots d^3p_{N_B} .
\end{equation}
We assume that $\rho(z,t)$ vanishes at infinity, so the last term in (\ref{contclasA}) is zero
by the Gauss theorem
\begin{equation}
 \int dz_B \nabla_B j_B \equiv \int_{\Gamma_B} dz_B \nabla_B j_B = 
 \int_{\partial \Gamma_B} dS_B\, j_B = 0,
\end{equation}
where $\partial \Gamma_B$ is the boundary at infinity of the infinite phase space $\Gamma_B$, while 
$dS_B$ is the surface area element on $\partial \Gamma_B$. Thus (\ref{contclasA}) can be written as a {\em new}
continuity equation
\begin{equation}\label{contclasAn}
 \frac{\partial\rho_A}{\partial t}+ \nabla_A j^{\rm tr}_A  =0 ,
\end{equation}
where 
\begin{equation}\label{rhoA}
 \rho_A(z_A,t)=\int dz_B\, \rho(z_A,z_B,t) ,
\end{equation}
\begin{equation}\label{jA}
 j^{\rm tr}_A(z_A,t)=\int dz_B\,j_A(z_A,z_B,t) .
\end{equation}
The quantity $\rho_A(z_A,t)$ defined by
(\ref{rhoA}) has the clear statistical interpretation; it is just the marginal probability density
of $z_A$, obtained by averaging over $z_B$. But what is $j^{\rm tr}_A(z_A,t)$ defined by (\ref{jA})?
The interpretation of $j^{\rm tr}_A(z_A,t)$ is a bit subtle, which we now analyze. 

First let us say that the label ``tr" stands for {\em truncated}, reminding us that 
$j^{\rm tr}_A(z_A,t)$ is a truncated version of $j_A(z_A,z_B,t)$, in the sense 
that $j^{\rm tr}_A(z_A,t)$ does not depend on $z_B$, 
i.e., the dependence on $z_B$ is truncated.\footnote{We also anticipate 
that in the quantum case the label ``tr" will acquire one additional 
meaning, namely ``traced", and both meanings will describe well the concept of $j^{\rm tr}_A$.}
The $\rho_A(z_A,t)$ is also truncated in that sense, but we do not use the notation
$\rho^{\rm tr}_A$ because the notation $\rho_A$ is already sufficient to distinguish it from
$\rho$.   

Writing (\ref{current}) as
\begin{equation}\label{currentA}
j_A=\rho v_A , \;\;\; j_B=\rho v_B , 
\end{equation}
and focusing on the $A$-part, we see that (\ref{jA}) can be written as 
\begin{equation}\label{jA2}
 j^{\rm tr}_A(z_A,t)=\int dz_B\,\rho(z_A,z_B,t)v_A(z_A,z_B,t) ,
\end{equation}
so $j^{\rm tr}_A(z_A,t)$ is an {\em average} of the current $j_A(z_A,z_B,t)$, where averaging is performed
over the whole $B$-subsystem. This motivates us to introduce the  
truncated velocity $v^{\rm tr}_A$ defined as
\begin{equation}
v^{\rm tr}_A(z_A,t)=\frac{j^{\rm tr}_A(z_A,t)}{\rho_A(z_A,t)}=\frac{\int dz_B\,j_A(z_A,z_B,t)}{\int dz_B\, \rho(z_A,z_B,t)} ,
\end{equation}
so that, in analogy with (\ref{currentA}), we can write
\begin{equation}
 j^{\rm tr}_A=\rho_A v^{\rm tr}_A ,
\end{equation}
where all the quantities depend only on $z_A$, not on $z_B$. Thus we see that the continuity equation 
(\ref{contclasAn}) is a natural continuity equation for the $A$-subsystem.

However, it is important to emphasize that the actual velocities of particles in the $A$-subsystem are 
given by $v_A$, not by $v^{\rm tr}_A$. The actual velocity is not the truncated velocity. In fact, a
truncated
velocity may be different from all possible actual velocities. There is, of course, 
nothing strange with it\footnote{It's not much different from the fact that an average family in a country may have 2.4 children, 
despite the obvious fact that no actual family has 2.4 children.}, 
when we keep in mind that truncation is a kind of averaging.

Even though the average velocities $v^{\rm tr}_A$ are not actual velocities 
of individual particles, in statistical physics 
it is not entirely wrong to think of them as actual velocities.
The actual velocities of individual particles are usually not measured in statistical physics, 
so no contradiction with experiments is produced by such thinking.
More importantly, such a way of thinking creates an intuitive picture 
associated with the continuity equation (\ref{contclasAn}) for the $A$-subsystem.
Creating such a picture with real velocities $v_A$ of the subsystem
is more complicated, because the real velocities, in general, 
obey a continuity equation only when the full closed system is taken into account.
But ultimately, the velocities $v^{\rm tr}_A$ are just a mathematical tool
in dealing with statistics of the $A$-subsystem.
The usual phase-space average value (to be distinguished from truncation average value)
of any observable $O(z_A)$ of the $A$-subsystem is given by
\begin{equation}\label{OA}
\bar{O}(t) = \int dz_A\, O(z_A)\rho_A(z_A,t) , 
\end{equation}
where $\rho_A(z_A,t)$ is the marginal probability density (\ref{rhoA}) which obeys the continuity equation
(\ref{contclasAn}). Ultimately, whether one interprets the velocities $v^{\rm tr}_A$ 
as real velocities of individual particles, or just as truncation average velocities,
does not influence the validity of the phase-space average (\ref{OA}), which is what we actually measure in statistical physics.  

Finally we note that, in general, the velocities $v^{\rm tr}_A$ are not described by Hamilton equations of motion. 
Thus, they cannot be treated by a formalism fully analogous to that in Sec.~\ref{SECclasclos}. 
In particular, there is no Liouville theorem for the continuity equation (\ref{contclasAn}).
Nevertheless, since the velocities $v^{\rm tr}_A$ obey the continuity equation (\ref{contclasAn}), 
they can be used as a basis for a consistent formulation of statistical mechanics.
 
\subsection{The physical meaning of phase-space average values}

In general, the phase-space average value such as (\ref{O}) or (\ref{OA}) has nothing to do with the actual value of the observable 
in the physical system. 
And yet, from an experimental point of view,
such an average value typically corresponds to a thermodynamic value that is measured in the laboratory.
This means that the average value in statistical physics can be interpreted as some sort of ``actual'' value,
at least in some approximative practical sense. What is the theoretical justification for such an interpretation?   
A short answer is that it is justified due to the law of large numbers, 
because in practice we usually assume that the number of particles $N$ or $N_A$ is large.  

Let us elaborate this a bit. When we consider an observable $O(z)=O(z_1,\cdots, z_N)$ for large $N$, we usually 
consider an observable which considers all $N$ particles on an equal footing. This means that the function 
$O(z_1,\cdots, z_N)$ is symmetric under all permutations of $z_a$'s. Hence, if $O(z)$ is an extensive quantity, such 
as the energy of the full system, the average value usually scales with $N$, i.e., we usually have $\bar{O}\propto N$.
Even if we study an intensive quantity, such as the pressure defined as force per unit area, or energy density defined as 
energy per unit volume, this is always applied to a relatively large area or volume containing a large number
$N$ of particles, so even for intensive quantities one deals with observables that usually scale with $N$. 
The standard deviation 
\begin{equation}
 \Delta O(t) = \sqrt{\overline{O^2}(t) - \bar{O}^2(t) } ,
\end{equation}
on the other hand, usually scales with $\sqrt{N}$, i.e., $\Delta O(t) \propto \sqrt{N}$.  
Hence the relative standard deviation scales as
 \begin{equation}
 \frac{\Delta O(t)}{\bar{O}(t)} \propto \frac{\sqrt{N}}{N} = \frac{1}{\sqrt{N}} ,
\end{equation}
which is negligible for large $N$. This is, in essence, why the average value in statistical physics 
of a large system can usually, in practice, be treated as the actual value.  

Essentially the same reasoning is valid also in quantum statistical physics.

\section{Quantum statistical mechanics in terms of probability currents}
\label{SECqstat}

\subsection{Quantum continuity equation of a closed system}
\label{SECqcc}

Similarly to Sec.~\ref{SECclasclos}, this subsection 
is a review of well known results in quantum mechanics, which are needed for a thorough understanding
of the rest of the paper.
%

Quantum mechanics is usually formulated in the Hilbert space, rather than the phase space. 
Thus, if we consider a closed system of $N$ particles without spin,
a convenient basis for the Hilbert space is the basis of position eigenstates 
$|x\rangle=|{\bf x}_1,\ldots,{\bf x}_N \rangle$, so an arbitrary time-dependent pure state $|\psi(t)\rangle$
is represented by the wave function
\begin{equation}
 \psi(x,t)=\langle x|\psi(t)\rangle .
\end{equation}
Assuming that it satisfies a Schr\"odinger equation of the form
\begin{equation}
 \hat{H}\psi(x,t)=i\hbar\partial_t \psi(x,t)
\end{equation}
with 
\begin{equation}
 \hat{H}=\sum_a \frac{-\hbar^2\mbox{\boldmath $\nabla$}_a^2}{2m_a} +V(x) ,
\end{equation}
the Schr\"odinger equation implies the continuity equation
\begin{equation}\label{cont}
 \frac{\partial\rho}{\partial t}+\sum_a \mbox{\boldmath $\nabla$}_a {\bf j}_a =0 ,
\end{equation}
where 
\begin{equation}\label{born}
 \rho(x,t)=\psi^*(x,t)\psi(x,t) ,
\end{equation}
\begin{eqnarray}\label{j_a}
 {\bf j}_a(x,t) &=& \frac{-i\hbar}{2m_a} [ \psi^*(x,t) \mbox{\boldmath $\nabla$}_a\psi(x,t) 
- (\mbox{\boldmath $\nabla$}_a\psi^*(x,t)) \psi(x,t)] 
\nonumber \\
 &\equiv& \frac{-i\hbar}{2m_a}  \psi^*(x,t) \!\stackrel{\leftrightarrow}{\mbox{\boldmath $\nabla$}}_a\! \psi(x,t) ,
\end{eqnarray}
are the probability density and the probability current, respectively. Similar expressions 
for the probability density and probability current exist also for 
more complicated Hamiltonians that may involve spin of the particle and/or electromagnetic vector potential 
${\bf A}({\bf x}_a,t)$, some of which we shall explicitly discuss later, but the important point is that there is always 
a continuity equation of the form (\ref{cont}). Since the continuity equation is the central equation in our approach,
we write it more compactly as 
\begin{equation}\label{contcomp}
 \frac{\partial\rho}{\partial t}+\nabla j =0, 
\end{equation}
which is fully analogous to the classical continuity equation (\ref{contclas2}), except that here
in quantum mechanics
it is the equation in the $N$-particle position space, rather than the $N$-particle phase space.
In this sense, the quantum continuity equation is even a bit simpler than the classical one.

Finally, we note that the continuity equation can also be written as 
\begin{equation}
 \frac{\partial\rho}{\partial t}+\nabla (\rho v) =0, 
\end{equation}
where 
\begin{equation}
 v\equiv \frac{j}{\rho} .
\end{equation}

\subsection{Quantum continuity equation of a subsystem}

Now we proceed completely analogously as in Sec.~\ref{SECclassub}, except that the phase-space points $z$ are replaced with 
position points $x$. We divide the full closed system into two subsystems $A$ and $B$, and write (\ref{contcomp}) as
\begin{equation}\label{contAB}
 \frac{\partial\rho}{\partial t}+\nabla_A j_A + \nabla_B j_B =0 .
\end{equation}
Then we integrate it over $dx_B$, which leads to the continuity equation of the $A$-subsystem
\begin{equation}\label{contAn}
 \frac{\partial\rho_A}{\partial t}+ \nabla_A j^{\rm tr}_A  =0 ,
\end{equation}
where 
\begin{equation}\label{rhoAq}
 \rho_A(x_A,t)=\int dx_B\, \rho(x_A,x_B,t) ,
\end{equation}
\begin{equation}\label{jAq}
 j^{\rm tr}_A(x_A,t)=\int dx_B\,j_A(x_A,x_B,t) ,
\end{equation}  
in full analogy with (\ref{contclasAn}), (\ref{rhoA}) and (\ref{jA}).

Note also that (\ref{contAB}) can be written in terms of velocities as
\begin{equation}\label{contABv}
 \frac{\partial\rho}{\partial t}+\nabla_A (\rho v_A) + \nabla_B (\rho v_B) =0 ,
\end{equation}
where
\begin{equation}
 v_A\equiv \frac{j_A}{\rho} ,\;\;\; v_B\equiv \frac{j_B}{\rho} .
\end{equation}
Likewise, (\ref{contAn}) can be written as
\begin{equation}\label{contAnv}
 \frac{\partial\rho_A}{\partial t}+ \nabla_A (\rho_Av^{\rm tr}_A) =0 ,
\end{equation}
where
\begin{equation}
 v^{\rm tr}_A \equiv \frac{j^{\rm tr}_A}{\rho_A}  .
\end{equation}

\subsection{Density operator and partial traces}

Many on the equations above can be written even more elegantly in terms of the density 
operator and partial traces. 

The state $|\psi(t)\rangle$ of the full closed system is represented 
by the density operator
\begin{equation}
 \hat{\rho}(t)=|\psi(t)\rangle\langle\psi(t)| .
\end{equation}
From the Schr\"odinger equation
\begin{equation}
 \hat{H}|\psi(t)\rangle =i\hbar\partial_t|\psi(t)\rangle 
\end{equation}
one derives the von Neumann equation
\begin{equation}\label{vN}  
\frac{\partial\hat{\rho}(t)}{\partial t} = -\frac{i}{\hbar}[\hat{H},\hat{\rho}(t)], 
\end{equation}
where $[\;,\;]$ denotes the commutator, which is analogous to the classical Liouville equation (\ref{Liouville}).

The matrix elements of the density operator in the $|x\rangle$ basis are
\begin{equation}\label{e48}
 \langle x|\hat{\rho}(t)|x'\rangle =\langle x|\psi(t)\rangle\langle\psi(t)|x'\rangle = \psi(x,t) \psi^*(x',t),
\end{equation} 
so 
from (\ref{born}) and (\ref{e48})
we see that
\begin{equation}\label{rhohatrho}
\rho(x,t)=\langle x|\hat{\rho}(t)|x\rangle .
\end{equation}
But note that $|x\rangle$ is not an eigenstate
of $\hat{H}$, so one cannot solve the von Neumann 
equation (\ref{vN}) only for the diagonal elements (\ref{rhohatrho}), without considering 
the whole density matrix $\langle x|\hat{\rho}(t)|x'\rangle$.
Hence, from the dynamical point of view, in principle, the whole density matrix is needed
to compute the diagonal elements (\ref{rhohatrho}).

A useful quantity is the partial traced density operator
\begin{equation}\label{tr1}
 \hat{\rho}_A(t)={\rm Tr}_B \hat{\rho}(t) = \int dx_B \langle x_B|\hat{\rho}(t)|x_B\rangle ,
\end{equation}
because its diagonal matrix elements  
\begin{eqnarray}\label{tr2}
 \langle x_A|\hat{\rho}_A(t)|x_A\rangle &=& \int dx_B \langle x_A,x_B|\hat{\rho}(t)|x_A,x_B\rangle
\nonumber \\
&=& \int dx_B\, \rho(x_A,x_B,t) = \rho_A(x_A,t) 
\end{eqnarray}
are nothing but (\ref{rhoAq}). Thus (\ref{rhoAq}) can be written in a rather elegant form
\begin{equation}\label{tr3}
 \rho_A(x_A,t)= \langle x_A| {\rm Tr}_B \hat{\rho}(t) |x_A\rangle .
\end{equation}

The currents can also be written in this language. First we introduce the velocity operator 
of the $a$'th particle
\begin{equation}\label{velop}
\hat{\bf v}_a=\frac{\hat{\bf p}_a}{m_a}= \frac{-i\hbar\mbox{\boldmath $\nabla$}_a}{m_a} ,
\end{equation}
so that (\ref{j_a}) can be written in several elegant forms as 
\begin{eqnarray}\label{eleg}
{\bf j}_a(x,t) &=& \Re\, \psi^*(x,t) \hat{\bf v}_a \psi(x,t) 
\nonumber \\
 &=& \Re\, \langle\psi(t)|x\rangle \langle x|\hat{\bf v}_a|\psi(t)\rangle  
= \Re\, \langle x|\hat{\bf v}_a|\psi(t)\rangle \langle\psi(t)|x\rangle
\nonumber \\
 &=& \Re\, \langle x| \hat{\bf v}_a \hat{\rho}(t) |x\rangle 
\nonumber \\
 &=& \langle x| \hat{\bf j}_a(t) |x\rangle ,
\end{eqnarray}
where $\Re$ denotes the real part and
\begin{equation}\label{curop}
  \hat{\bf j}_a(t) \equiv \frac{{\hat{\bf v}_a\hat{\rho}(t)+\hat{\rho}(t)\hat{\bf v}_a}}{2}
= \frac{ \{ \hat{\bf v}_a,\hat{\rho}(t) \} }{2}
\end{equation}
is the probability current operator of the $a$'th particle, while $\{\; ,\;\}$ denotes the anti-commutator.
Then using the compact notation 
$\hat{j}=(\hat{\bf j}_1,\ldots,\hat{\bf j}_N)$, we can write (\ref{eleg}) as
\begin{equation}
 j(x,t)=\langle x|\hat{j}(t) |x\rangle ,
\end{equation}
which has the same form as (\ref{rhohatrho}). Thus, in analogy with (\ref{tr1})-(\ref{tr3}), 
we readily get
\begin{equation}\label{tjA}
 j^{\rm tr}_A(x_A,t)= \langle x_A| \hat{j}^{\rm tr}_A(t) |x_A\rangle ,
\end{equation}
where
\begin{equation}\label{hattilA}
\hat{j}^{\rm tr}_A(t) = {\rm Tr}_B \hat{j}_A(t) . 
\end{equation}
From (\ref{tjA}) we see that the label ``tr" can be interpreted as ``truncated" in the same sense as in the classical case, but from (\ref{hattilA}) we see that it can also be interpreted as
``traced". In the quantum case, both interpretations describe well the meaning of ``tr". 
In any case, it should be kept in mind that $\hat{j}^{\rm tr}_A(t)$ is different from $\hat{j}_A(t)$ in the decomposition
\begin{equation}
 \hat{j}(t) =\left( \hat{j}_A(t), \hat{j}_B(t) \right) .
\end{equation}
They, in fact, live in different Hilbert spaces.
The current operator $\hat{j}_A(t)$ is an operator in the full Hilbert space of the closed system,
while $\hat{j}^{\rm tr}_A(t)$ is an operator in the Hilbert space of the $A$-subsystem. 

The truncated current operator (\ref{hattilA}) of the $A$-subsystem can also be written in a form similar to (\ref{curop}).
With the notation $\hat{v}=(\hat{\bf v}_1,\ldots,\hat{\bf v}_N) = (\hat{v}_A,\hat{v}_B)$,
(\ref{curop}) can be written as 
\begin{equation}\label{curopAB}
  \hat{j}_A(t) = \frac{ \{ \hat{v}_A,\hat{\rho}(t) \} }{2} , \;\;\;
\hat{j}_B(t) = \frac{ \{ \hat{v}_B,\hat{\rho}(t) \} }{2} .
\end{equation}
Focusing on the $A$-part, (\ref{hattilA}) is  
\begin{equation}\label{hattilA2}
\hat{j}^{\rm tr}_A(t) = \frac{ {\rm Tr}_B \{ \hat{v}_A,\hat{\rho}(t) \} }{2}
= \frac{ {\rm Tr}_B [\hat{v}_A\hat{\rho}(t)] + {\rm Tr}_B [\hat{\rho}(t)\hat{v}_A] }{2} .
\end{equation}
But the operator $\hat{v}_A$ acts non-trivially
only in the Hilbert space of the $A$-subsystem. More precisely, 
if we write the full Hilbert space as ${\cal H}={\cal H}_A\otimes{\cal H}_B$,
with a little abuse of notation we can write $\hat{v}_A=\hat{v}_A \otimes 1$.  
Hence
\begin{equation}
{\rm Tr}_B [\hat{v}_A\hat{\rho}(t)]=\hat{v}_A{\rm Tr}_B\hat{\rho}(t) =\hat{v}_A \hat{\rho}_A(t) .
\end{equation}
(Recall that the velocity operators (\ref{velop})
are abstract operators defined by their action on the set of {\em all} differentiable functions,
so they do not depend on a particular position-dependent wave function, implying 
that the operators do not depend on the particle positions. The velocity operators should be distinguished 
from Bohmian velocities to be discussed later, which depend on the particular wave function and hence on the positions.)
Similarly ${\rm Tr}_B [\hat{\rho}(t) \hat{v}_A]=\hat{\rho}_A(t) \hat{v}_A$, 
so (\ref{hattilA2}) simplifies to
\begin{equation}\label{hattilA3}
\hat{j}^{\rm tr}_A(t) = \frac{ \hat{v}_A \hat{\rho}_A(t) + \hat{\rho}_A(t) \hat{v}_A}{2} 
= \frac{ \{ \hat{v}_A, \hat{\rho}_A(t) \} }{2} ,
\end{equation}
which has the form similar to the first equation in (\ref{curopAB}). 
Thus, if we know the reduced (i.e. truncated) density operator $\hat{\rho}_A(t)$, 
then the truncated current operator  
$\hat{j}^{\rm tr}_A(t)$ is expressed in terms of $\hat{\rho}_A(t)$ by (\ref{hattilA3}).

The fact that the truncated density operator truly describes a subsystem is justified by the
facts that it fulfills all requirements of a density operator, and that the ensemble average of an
observable of a subsystem is computed by the truncated density operator. 
The truncated probability current is justified as a valid description of a subsystem 
because it obeys a continuity equation associated with the truncated density,
thus providing that the sum of all probabilities of positions in a subsystem is equal to one at all times.

\subsection{Spin}

So far we were considering particles without spin. Now we shall describe how the formalism developed above 
can be generalized to include the spin.

Consider first a single particle with spin-$\frac{1}{2}$. In non-relativistic quantum mechanics it can be described 
by a 2-component wave function
\begin{equation}
 \psi({\bf x},t)=
\begin{pmatrix}
\psi_1({\bf x},t) \\
\psi_2({\bf x},t) 
\end{pmatrix}
.
\end{equation}
The corresponding probability density of particle positions is then
\begin{equation}
 \rho({\bf x},t)=\psi^{\dagger}({\bf x},t)\psi({\bf x},t)=\sum_{s=1,2} \psi^*_s({\bf x},t)\psi_s({\bf x},t) ,
\end{equation}
where $s$ is the spin index. It is illuminating to introduce the notation
\begin{equation}
\psi_s({\bf x},t)\equiv\psi(s,{\bf x},t)=\langle s,{\bf x}|\psi(t)\rangle, 
\end{equation}
so that we can write
\begin{eqnarray}
 \rho({\bf x},t) &=& \sum_s \psi^*(s,{\bf x},t)\psi(s,{\bf x},t) 
\nonumber \\
&=& \sum_s  \langle\psi(t)|s,{\bf x}\rangle  \langle s,{\bf x}|\psi(t)\rangle
= \sum_s \langle s,{\bf x}|\psi(t)\rangle \langle\psi(t)|s,{\bf x}\rangle
\nonumber \\
&=& \sum_s \langle s,{\bf x}|\hat{\rho}(t)|s,{\bf x}\rangle
= \langle{\bf x}|{\rm Tr}_S\hat{\rho}(t)|{\bf x}\rangle ,
\end{eqnarray}
where
\begin{equation}
 {\rm Tr}_S\hat{\rho}=\sum_s \langle s|\hat{\rho}|s\rangle 
\end{equation}
denotes the partial trace in the spin part of the Hilbert space.

Next let us generalize it to $N$ particles with spin. The wave function has $N$ spinor indices
\begin{equation}
 \psi_{s_1\ldots s_N}({\bf x}_1,\ldots, {\bf x}_N,t)=\psi_s(x,t) ,
\end{equation}
where $s=(s_1,\ldots, s_N)$ is a compact notation denoting a collective spin index.  Thus the formula 
above for the probability density in the position space readily generalizes to
\begin{equation}
 \rho(x,t)=\sum_{s} \psi^*_s(x,t)\psi_s(x,t)=\langle x|{\rm Tr}_S\hat{\rho}(t)|x\rangle .
\end{equation}

The probability current for the particles with spin can be written down analogously.
It is simply
\begin{eqnarray}
 j(x,t) &=& \Re\sum_{s} \psi^*_s(x,t)\hat{v}\psi_s(x,t)
\nonumber \\
&=& \langle x|{\rm Tr}_S\hat{j}(t)|x\rangle .
\end{eqnarray} 

Finally, the probability density and the current for the $A$-subsystem are the generalizations of 
(\ref{tr3}) and (\ref{tjA})
\begin{eqnarray}\label{tr3S}
 \rho_A(x_A,t) &=& \langle x_A| {\rm Tr}_{S_A} {\rm Tr}_B\, \hat{\rho}(t) |x_A\rangle 
\nonumber \\
&=& \langle x_A| {\rm Tr}_{S_A} \hat{\rho}_A(t) |x_A\rangle ,
\end{eqnarray}
\begin{eqnarray}\label{tjAS}
 j^{\rm tr}_A(x_A,t) &=& \langle x_A| {\rm Tr}_{S_A} {\rm Tr}_B\, \hat{j}_A(t) |x_A\rangle 
\nonumber \\
&=& \langle x_A| {\rm Tr}_{S_A} \hat{j}^{\rm tr}_A(t) |x_A\rangle .
\end{eqnarray}
Here ${\rm Tr}_{S_A}$ is the trace over the spin space of the $A$-subsystem, 
${\rm Tr}_B={\rm Tr}_{S_B} {\rm Tr}_{X_B}$ is the trace over all space of the $B$-subsystem,
and ${\rm Tr}_{X_B}$ is the trace over the position space of the $B$-subsystem.

When the particles are identical, the wave function of the whole system is either symmetric (for integer spins)
or antisymmetric (for half-integer spins) under the exchange of all particles, so the probability density and the current 
are symmetric. This provides that the truncated probability density and current of the subsystem are symmetric too. 

\subsection{Summary}

Now let us summarize the main results of this rather lengthy and formal section,
and introduce one additional notation for partial traces that further illuminates the whole idea. 

A closed system of $N$ particles is 
described by the pure state $|\psi(t)\rangle$, corresponding to the density operator 
\begin{equation}
\hat{\rho}(t)=|\psi(t)\rangle\langle\psi(t)|.
\end{equation}
Introducing the $N$-particle probability current operator
\begin{equation}
 \hat{j}(t)=\frac{\hat{v}\hat{\rho}(t) + \hat{\rho}(t)\hat{v}}{2} ,
\end{equation}
where $\hat{v}$ is the $N$-particle velocity operator, one defines the probability density and the $N$-particle 
probability current as
\begin{equation}\label{rho1}
 \rho(x,t)=\langle x|{\rm Tr}_{({\rm no}\;X)} \hat{\rho}(t)|x\rangle ,
\end{equation}
\begin{equation}\label{j1}
 j(x,t)=\langle x|{\rm Tr}_{({\rm no}\;X)} \hat{j}(t)|x\rangle ,
\end{equation}
where ${\rm Tr}_{({\rm no}\;X)}$ denotes the partial trace over everything 
which is {\em not} the $X$-space (the Hilbert space spanned by the position basis vectors $|x\rangle$).

Then we split the closed system into two subsystems $A$ and $B$, having $N_A$ and $N_B$ particles, 
respectively, so that $N_A+N_B=N$. The full current operator can be decomposed as
\begin{equation}
 \hat{j}(t)= \left( \hat{j}_A(t), \hat{j}_B(t) \right) .
\end{equation}
Focusing on the $A$-part, the probability density and the $N_A$-particle probability current for the $A$-subsystem are
\begin{eqnarray}\label{rho2}
 \rho_A(x_A,t) &=& \langle x_A|{\rm Tr}_{({\rm no}\;X_A)} \hat{\rho}(t)|x_A\rangle
\nonumber \\ 
 &=& \langle x_A|{\rm Tr}_{S_A} \hat{\rho}_A(t)|x_A\rangle ,
\end{eqnarray}
\begin{eqnarray}\label{j2}
 j^{\rm tr}_A(x_A,t) &=& \langle x_A|{\rm Tr}_{({\rm no}\;X_A)} \hat{j}_A(t)|x_A\rangle 
\nonumber \\ 
 &=& \langle x_A| {\rm Tr}_{S_A} \hat{j}^{\rm tr}_A(t)|x_A\rangle ,
\end{eqnarray}
where 
\begin{equation}
 \hat{\rho}_A(t) = {\rm Tr}_B \hat{\rho}(t) ,
\end{equation}
\begin{eqnarray}\label{sumtj}
 \hat{j}^{\rm tr}_A(t) &=& {\rm Tr}_B \hat{j}_A(t)
\nonumber \\ 
&=&  \frac{ \hat{v}_A \hat{\rho}_A(t) + \hat{\rho}_A(t) \hat{v}_A }{2} .
\end{eqnarray}

The density (\ref{rho1}) and current (\ref{j1}) satisfy the continuity equation
\begin{equation}\label{contfin}
 \frac{\partial\rho}{\partial t}+\nabla j=0 .
\end{equation}
Likewise, the density (\ref{rho2}) and current (\ref{j2}) satisfy the continuity equation
\begin{equation}\label{contAfin}
 \frac{\partial\rho_A}{\partial t}+\nabla_A j^{\rm tr}_A=0 .
\end{equation}

\section{Bohmian interpretation}
\label{SECbohm}

\subsection{Bohmian interpretation of the closed system}
\label{SECbc}

Similarly to Sec.~\ref{SECclasclos} and Sec.~\ref{SECqcc}, this subsection 
is a review of the results already known in the literature  \cite{bohm,book-bohm,book-hol,book-durr,oriols}.

The Bohmian interpretation offers an intuitive interpretation of the quantum continuity equation (\ref{contfin}).
Writing it as
\begin{equation}\label{contB}
 \frac{\partial\rho}{\partial t}+\nabla (\rho v)=0 ,
\end{equation}
where 
\begin{equation}\label{vB}
 v=\frac{j}{\rho} ,
\end{equation}
the Bohmian interpretation postulates that each particle in the full closed system has an actual trajectory
${\bf x}_a(t)$, $a=1,\ldots, N$,
satisfying the equation of motion $d{\bf x}_a/dt={\bf v}_a$. 
This is similar to the classical equation of motion (\ref{eomclas}), but in the Bohmian 
interpretation we are working in the position space, not the phase space.
In our compact notation, the Bohmian equation of motion for particle trajectories $x(t)$ is just
\begin{equation}\label{traj}
 \frac{dx}{dt}=v .
\end{equation}

According to the Bohmian interpretation, the world is made of pointlike particles having trajectories guided by the law 
(\ref{traj}). The quantum velocity function $v(x,t)$ is analogous to the classical velocity function 
$v(z)=({\bf v}_1(z),\ldots ,{\bf v}_n(z))$ defined by (\ref{va}). In particular, just as classical $v(z)$ 
is computed from the Hamiltonian $H(z)$, the Bohmian $v(x,t)$ is computed from the wave function $\psi_s(x,t)$, 
or equivalently, from the state in the Hilbert space $|\psi(t)\rangle$, 
i.e., from the density operator $\hat{\rho}(t)=|\psi(t)\rangle\langle\psi(t)|$. In this sense, the density operator 
$\hat{\rho}(t)$ in Bohmian mechanics plays a role similar to the Hamiltonian $H(z)$ in classical mechanics.
According to the Bohmian interpretation, 
the $N$-particle trajectory $x(t)$ represents the {\em ontology} of the world, namely the things the world is made of,    
while $\hat{\rho}(t)$ represents the {\em nomology} 
\cite{nomological}
of the world, namely the object that concisely encodes 
the law governing the trajectory $x(t)$. 

Even though the Bohmian law of motion (\ref{traj}) is deterministic, it is compatible with the usual probabilistic nature 
of quantum mechanics (QM). Namely, the probabilities arise in the same was as in classical statistical mechanics, 
through the lack of knowledge of actual initial conditions. Assuming that the initial probability density of initial 
particle positions in a statistical ensemble is given by $\rho(x,t_0)$, 
the continuity equation (\ref{contB}) and particle velocities (\ref{traj}) provide that the probability density of particle 
positions is equal to $\rho(x,t)$ at {\em any} time $t$. This is how
Bohmian mechanics makes the same probabilistic predictions in the $x$-space as standard QM in the $x$-space.

But what about other spaces? QM also makes probabilistic predictions in the momentum space, energy space,
spin space, etc. How can Bohmian mechanics make {\em any} predictions in these other spaces?
And how can these Bohmian predictions be compatible with the predictions of standard QM?

To answer this question, the key word is {\em measurement}. For {\em any} quantum observable 
(position, momentum, energy, spin, ...),
Bohmian mechanics predicts the same probabilities as standard QM {\em when this observable is measured}.
But how? Essentially, by reducing measurement of {\em any} observable to an observation of a 
{\em position} of something, and using the fact that position probabilities in Bohmian mechanics are the same 
as position probabilities in standard QM.
But how can measurement of {\em any} observable be reduced to an observation of position?
In general, measurement is a physical process which establishes a correlation between the measured object 
and some {\em macroscopic pointer of the measuring apparatus}. But in practice, 
the pointer of the measuring apparatus can always be described in terms of a macroscopic {\em position},
so in practice any measurement is really an observation of a macroscopic position. 
And of course, the macroscopic position of the pointer is just an aggregate of all the microscopic Bohmian
positions $x_P$ of the pointer $P$.  
In a nutshell, this is how Bohmian mechanics explains the results of all quantum measurements as observations of positions.
For more details see e.g. \cite{bohm,book-bohm,book-hol,book-durr,oriols,nikIBM}.      

Since the world, according to the Bohmian interpretation, is made of pointlike particles obeying the equation of motion
(\ref{traj}), it follows that we only need the continuity equation (\ref{contB}) in the $x$-space. 
There is no need for a continuity equation in other spaces, such as momentum space or spin space.
Nevertheless, it is worth mentioning that Bohmian mechanics can be extended to the momentum space as well \cite{bonilla20},
and that Bohmian trajectories can be interpreted within standard quantum mechanics as
borders of regions of constant probability \cite{bonilla22}.

\subsection{Bohmian interpretation of a subsystem}

In principle, the Bohmian interpretation of the $A$-subsystem is straightforward. From (\ref{traj})
it follows that particles of the $A$-subsystem obey the equation of motion
\begin{equation}\label{trajA}
 \frac{dx_A}{dt}=v_A .
\end{equation}
However, the right-hand side $v_A(x,t)=v_A(x_A,x_B,t)$ depends not only on the positions $x_A$ in the $A$-subsystem,
but also on the positions $x_B$ in the $B$-subsystem. Thus, strictly speaking, we cannot study 
a subsystem without considering the whole system. In principle, any subsystem is more or less entangled 
with the rest of the Universe. In Bohmian mechanics, it reflects in the fact that particle velocities
in any subsystem depend on particle positions in the whole closed system. 

Nevertheless, there is a way to apply Bohmian mechanics to a subsystem {\em without} considering the rest of the system.
The idea is to use the continuity equation (\ref{contAfin}) written as 
 \begin{equation}\label{contA_B}
 \frac{\partial\rho_A}{\partial t}+\nabla_A (\rho_Av^{\rm tr}_A)=0 ,
\end{equation}
where 
\begin{equation}\label{vB_A}
 v^{\rm tr}_A=\frac{j^{\rm tr}_A}{\rho_A} .
\end{equation}
Thus, if one assumes that particles of the $A$-system have trajectories $x^{\rm tr}_A(t)$ obeying
\begin{equation}\label{trajAt}
 \frac{dx^{\rm tr}_A}{dt}=v^{\rm tr}_A ,
\end{equation}
then one can explain all the probabilistic predictions of standard QM in the $A$-subsystem, in a way completely analogous
to that Sec~\ref{SECbc}, provided that the $A$-subsystem is macroscopically large.
Namely, if the $A$-subsystem is macroscopically large, then the aggregates of its particles 
can form macroscopic objects, such as tables, chairs, gases, and, of course, macroscopic measuring apparatuses.
Since $v^{\rm tr}_A(x^{\rm tr}_A,t)$ on the right hand-side of (\ref{trajAt}) 
does not depend on $x_B$, it follows that truncated trajectories obeying (\ref{trajAt}) do not depend on particle positions 
of the $B$-subsystem. This makes the explanation based on (\ref{trajAt}) simpler than the 
explanation based on (\ref{trajA}), at least if simplicity is counted by the number of particles 
that one must take into account. 

Of course, since we assumed that the correct equation of motion is (\ref{trajA}), it follows that 
(\ref{trajAt}) is strictly speaking ``wrong''. Nevertheless, in practice, we never observe 
the microscopic Bohmian trajectories, we only observe their macroscopic aggregates. 
This means that, for all practical purposes, the trajectories (\ref{trajA}) make the same 
measurable predictions as the trajectories (\ref{trajAt}). As long as one is only interested in the $A$-subsystem,
for all practical purposes the trajectories (\ref{trajAt}) defined by the truncated velocities
$v^{\rm tr}_A$ are not any less right than the trajectories (\ref{trajA}) defined by the velocities
$v_A$.

\subsection{Deeper meaning of the Bohmian interpretation}

Many adherents of the Bohmian interpretation like to think that the purpose of Bohmian mechanics
is to tell us what is {\em the} right ontology of the world. While we fully sympathize with 
such a point of view, in our opinion this is not the best way, and certainly not the only way,
to understand the Bohmian interpretation \cite{nikIBM}. All theories in physics are just provisional 
theories that one day may be replaced by better theories. Or to use the language common in 
quantum field theory \cite{weinberg}, all theories that we have are just effective theories, 
applicable in some regime, but not in all regimes. For example, in this paper we study 
systems with a fixed number $N$ of particles in the closed system, so this theory is not applicable 
in the regime where particles are created and destroyed. The creation and destruction of particles 
is described by relativistic quantum field theory (QFT), which we do not discuss in this paper
(see e.g. \cite{nikQFTproof_concept} for a light discussion of Bohmian formulation(s) of relativistic QFT). Even 
relativistic QFT, as we currently understand it, is probably not applicable at short distances comparable to the 
Planck distance, where effects of quantum gravity are expected to become important. 
Hence, instead of saying that Bohmian mechanics tells us {\em the} right ontology of the world,
it is much more sober to say that Bohmian mechanics tells us {\em a} right ontology of the world.
Namely, this ontology is ``right'' only in an effective sense, because it provides a useful theoretical tool
that helps in intuitive thinking about a large class of quantum phenomena. 
But it is not ``right'' in an absolute sense, because a better more fundamental theory may be much more easily 
understood in terms of an entirely different ontology. Ontology, just like any other concept in theoretical physics,
is nothing but a useful\footnote{The notion of ``useful'' is of course subjective. What is useful for one person
may not be useful for another. In particular, there are many interpretations of QM,
and most physicists don't see the Bohmian interpretation as particularly useful. 
Those who don't find it useful, should not use it. Those who do find it useful, should use it.
The Bohmian interpretation, or any other interpretation for that matter, is just a thinking tool.
The tools are not right or wrong. The tools are useful or useless. And this is subjective.}
thinking tool. A useful fiction, if you like.
This is the basis of the instrumental interpretation of Bohmian mechanics, 
advocated also in \cite{nikIBM}.    

From this point of view, neither the trajectories (\ref{traj}) nor the trajectories (\ref{trajAt})
are ``right'' in an absolute sense. Both are just effective theories, describing ontologies
suitable for simple intuitive descriptions of the systems they study. 
The trajectories (\ref{traj}) represent a simple ontology suitable for a {\em closed} system 
of a fixed number $N$ of non-relativistic particles. The truncated trajectories (\ref{trajAt}) 
represent a simple ontology 
suitable for an {\em open} subsystem of a fixed number $N_A$ of non-relativistic particles. 
The trajectories (\ref{traj}) are more powerful than the truncated trajectories (\ref{trajAt}), because 
(\ref{traj}) can describe both the full closed system {\em and} the open $A$-subsystem satisfying (\ref{trajA}). 
Nevertheless, the description of the $A$-subsystem
in terms of (\ref{trajAt}) is simpler. Thus the truncated theory (\ref{trajAt}) can be thought of as effective theory for the more 
fundamental theory (\ref{traj}). But ultimately both are just effective theories for some even more 
fundamental theory.  

Such an effective view of the Bohmian interpretation allows a lot of ontological flexibility. 
Depending on the context, one can change which entities will be considered ontological, and which 
entities will not be considered so. The formalism developed in Sec.~\ref{SECqstat}
provides an elegant, systematic and powerful way to translate this choice into a mathematical language.
In essence, all things which are not considered ontological are traced out in the density operator 
$\hat{\rho}(t)$. In the Bohmian interpretation spin is usually not considered ontological, 
which is encoded in the partial trace ${\rm Tr}_S$ over the spin part of the Hilbert space. 
Likewise, if one does not want to treat the $B$-subsystem as ontological, one performs 
the partial trace ${\rm Tr}_B$. In principle, one can trace over everything except the things 
one is really interested about in a given context. For example, if one only wants to know 
how the macroscopic pointer of the measuring apparatus behaves, and does not care about how the 
microscopic measured system behaves, one can trace over everything except the pointer degrees of freedom.
This may seem like an extreme application of partial tracing, but such an extremely 
truncated version of Bohmian interpretation also leads to the same measurable predictions 
as standard QM \cite{nik_solip}. An even more extreme version is tracing over everything except the relevant parts 
of the brain of the conscious observer, which again makes the same observable predictions 
as standard QM \cite{nik_solip}. Such {\it ad hoc} truncations of ontology may look crazy from a philosophical point of view,
but from our practical effective ontology point of view there is nothing wrong with them, as long as 
one finds them useful as thinking tools.

In addition to giving the ontology, it has been argued that Bohmian mechanics also makes a prediction for the arrival probability 
density in terms of the probability current \cite{leavens}. But the same prediction for the arrival probability density can also 
be obtained from standard quantum mechanics \cite{nik_arr1,nik_arr2}, without reference to Bohmian mechanics, so the prediction 
does not really depend on whether the Bohmian trajectories are ontological. The continuity equation is one of the crucial ingredients
in the methods developed in \cite{nik_arr1,nik_arr2}, so these methods  
can straightforwardly be extended to open subsystems to associate arrival probability density with the truncated current.

\section{Explanation of standard statistical mechanics}
\label{SECexplssm}

\subsection{Motivation}

The main purpose of statistical mechanics is to give a simplified description of large systems, typically 
containing $10^{23}$ or more particles. Obviously, in practice, one cannot give a detailed description 
of each and every particle in such a large system, so one needs a practical approximation 
that describes the collective behavior of the large system without a detailed description of every particle.
Statistical mechanics can also be applied to small systems, but that is not its main purpose. 
Unless explicitly stated otherwise, in this section we shall always tacitly assume that the system 
of interest is large. 

In quantum statistical mechanics, the system of interest is usually described by a mixed density operator $\hat{\rho}$.
In particular, for a system with a fixed number of particles in thermal equilibrium, the density operator is
\begin{equation}\label{Srho}
 \hat{\rho}=\frac{e^{-\beta \hat{H}}}{ {\rm Tr}\, e^{-\beta \hat{H}} } ,
\end{equation}
where $\beta=1/(kT)$, $k$ is the Boltzmann constant, $T$ is the temperature, and $\hat{H}$ is the Hamiltonian of the system.
With the aid of $\hat{\rho}$ one can compute the average value of any observable $\hat{O}$ one is interested about, by the formula
\begin{equation}\label{qaverage}
 \bar{O}={\rm Tr} (\hat{O}\hat{\rho}) .
\end{equation}
For practical purposes, this is often more-or-less all that one needs to know.
In this paper, however, we do not deal with practical problems as such. Instead, our goal 
is to better understand quantum statistical mechanics from a foundational conceptual point of view.

One conceptual question is the meaning of the mixed state $\hat{\rho}$. For instance,
(\ref{Srho}) can be written as 
\begin{equation}\label{Srho2}
 \hat{\rho} \propto \sum_n e^{-\beta E_n } |n\rangle\langle n|,
\end{equation}
where $|n\rangle$ are eigenstates of $\hat{H}$, $\hat{H}|n\rangle = E_n|n\rangle$,
and, for simplicity, we have ignored the normalization constant $( {\rm Tr}\, e^{-\beta \hat{H}} )^{-1}$.
Does it mean that the system is actually in only one of the energy eigenstates $|n\rangle$, but we just don't know which one?
Or does it mean that the system actually {\em is} in the mixed state (\ref{Srho2}), irrespective of our knowledge?
Or is this question just meaningless, because the average value (\ref{qaverage}) does not depend on it?

Standard practical textbooks usually do not discuss such conceptual questions, 
to address such questions one needs to go deeper into the field 
of foundations of quantum statistical mechanics. 
In this section we shall study how such foundational questions can be answered with the aid of Bohmian interpretation.
Of course, such questions can also be studied without the Bohmian interpretation, and indeed, a large part of this section 
will not depend on the Bohmian interpretation. However, among all interpretations of QM, the Bohmian interpretation is 
the most ``classical'', in the sense that it entails the same ontology as classical mechanics.
Hence, since quantum statistical mechanics usually deals with macroscopic systems, i.e., systems
which show both classical and quantum features, we find the Bohmian interpretation particularly convenient.

\subsection{Ontology {\it vs} nomology}

In Sec.~\ref{SECbc} we made a distinction between ontology and nomology. Here we want to elaborate this distinction 
in more detail.

There is no precise definition of the notions of ontology and nomology, 
yet their distinction is useful.\footnote{There are many concepts in physics that are not precisely defined.
Some examples are microscopic, macroscopic, fundamental and emergent. 
Even though they lack a precise definition, they are a part of normal physics vocabulary,
because they are useful in conceptual thinking.}
The easiest way to absorb their meaning is through examples.

Consider first classical mechanics of pointlike particles. The world, according to this theory, is made of particles 
with trajectories $x(t)$. We say that the trajectories $x(t)$ are ontological. Anything directly derivable 
from $x(t)$, like velocities $dx(t)/dt$, is also ontological. The nomology, on the other hand, is the set of physical
{\em laws} that govern the behavior of the ontological stuff. Hence the mathematical objects in the theory 
that encode the physical laws are called nomological. For instance, the Hamiltonian and the Lagrangian are 
nomological. What about energy? The energy depends on both the ontological stuff $x(t)$ and $dx(t)/dt$, {\em and} 
the nomological Hamiltonian, so energy is partially ontological and partially nomological.   
The masses $m_a$ of the particles, which are parameters in the Hamiltonian, are nomological.
The particle momenta $p(t)=({\bf p}_1(t), \ldots, {\bf p}_N(t))$, ${\bf p}_a(t)=m_a d{\bf x}_a(t)/dt$, 
are hence partially ontological and partially nomological. 

Now consider Bohmian mechanics. It has classical ontology defined by trajectories $x(t)$, but the laws 
that govern $x(t)$ are very different from the classical laws. Classical and Bohmian mechanics have the same 
ontology but different nomology. As we already explained in Sec.~\ref{SECbc}, the nomology of Bohmian 
mechanics can be concisely encoded in the density operator $\hat{\rho}(t)$. The density operator itself
is governed by the Hamilton operator $\hat{H}$, so $\hat{H}$ is also nomological.
But the Bohmian law of particle motion does not have a form of Hamilton equations of motion, so energy 
of particles in Bohmian mechanics is not a well defined quantity. In this sense, the energy is not even partially 
ontological in Bohmian mechanics.

Nevertheless, in quantum statistical mechanics, one can define energy as the average energy 
\begin{equation}\label{Eaverage}
 E(t)\equiv\bar{E}(t)={\rm Tr}\, \hat{H}\hat{\rho}(t) .
\end{equation}
Here $\hat{\rho}(t)$ can be either a pure state of a closed system or a mixed state
of an open system, which is why, in general, $E(t)$ can depend on time $t$.
Clearly, this energy is {\em purely nomological}. The ``average'' in (\ref{Eaverage})  
is a formal abstract quantum average, which does not involve any averaging over ontological particle positions.

\subsection{Different kinds of entropy}
\label{SECentropy}

The simplest definition of entropy in quantum statistical mechanics is von Neumann entropy 
\begin{equation}\label{vNe}
S_{\rm vN}(t)=-k {\rm Tr}\, \hat{\rho}(t)\ln \hat{\rho}(t) . 
\end{equation}
Similarly to the energy (\ref{Eaverage}), this entropy
is purely nomological. For a closed system it does not depend on time,
due to the von Neumann equation (\ref{vN}).
For an open system this is really the entanglement entropy, because 
$\hat{\rho}(t)$ is a mixed state due to entanglement with the rest of the Universe.
Hence, in open systems, this entropy tends to increase with time, because, if initially 
the open system is not maximally entangled with the rest of the Universe, the interaction with the 
rest of the Universe usually creates more entanglement, which increases the entanglement entropy. 

Another useful definition of entropy is quantum Boltzmann entropy  
\begin{equation}\label{SqB}
 S_{\rm qB}({\cal H})=k \ln\dim {\cal H} ,
\end{equation}
where $\dim {\cal H}$ is the dimension of a Hilbert space ${\cal H}$.
To be meaningful, the ${\cal H}$ must be a finite-dimensional Hilbert space.  
Thus ${\cal H}$ is not the whole Hilbert space, which in quantum mechanics is usually infinite dimensional,
but a finite dimensional subspace. For example, the Hilbert space of wave functions for a single free particle in the infinite 
3-dimensional space is infinite dimensional, but a finite dimensional Hilbert subspace is obtained if we only 
consider wave functions confined inside a finite volume $V$, and discard all wave functions corresponding to the momentum 
larger than some arbitrary cutoff momentum $p_{\rm cutoff}$. 
In practice, ${\cal H}$ is usually chosen such that all different states in ${\cal H}$
cannot be distinguished at the macroscopic level. This, of course, depends on how  
``distinguished at the macroscopic level'' is defined. In practice, this us usually 
defined intuitively. More fundamentally, the ability to macroscopically distinguish things depends 
on the Hamiltonian $\hat{H}_O$ that describes the observer. 
In any case, the quantum Boltzmann entropy (\ref{SqB}) is nomological.

Quantum Boltzmann entropy becomes particularly interesting if one decomposes the full Hilbert space
${\cal H}$ into a direct sum of finite dimensional Hilbert spaces 
\begin{equation}\label{oplus}
 {\cal H} = \bigoplus_M {\cal H}_M ,
\end{equation}
where each ${\cal H}_M$ corresponds to one macroscopic state that can be distinguished from other 
macroscopic states. Then, for any pure state $|\psi\rangle$ living in only one subspace ${\cal H}_M$,
i.e., for any $|\psi\rangle$ obeying $|\psi\rangle \in {\cal H}_M$,
one can define quantum Boltzmann entropy as 
\begin{equation}\label{qBM}
 S_{\rm qB}(|\psi\rangle)=k \ln\dim {\cal H}_M(|\psi\rangle) .
\end{equation} 
This defines a non-zero entropy for a pure state. 
Moreover, if there is a mechanism by which $|\psi(t)\rangle$ at any time $t$ lives in only one subspace ${\cal H}_{M(t)}$,
then the entropy (\ref{qBM}) changes with time. If it starts in a low entropy state, 
such a time-dependent quantum Boltzmann entropy naturally tends to increase with time,
because the prior probability that such a state is an element of ${\cal H}_M$ is proportional to $\dim {\cal H}_M$, 
so growth of entropy is just a natural transition to a more probable macroscopic state.     
Clearly, the quantum Boltzmann entropy (\ref{qBM}) is also nomological.

The quantum Boltzmann entropy can also be viewed as a function of the ontological 
Bohmian position $x(t)$. In practice, macroscopically distinguishable states are 
distinguishable in the $x$-space, which means that wave functions from different ${\cal H}_M$'s 
have a negligible overlap in the $x$-space. Hence, at almost any time $t$,
$x(t)$ can be considered to be in only one ${\cal H}_M$, so we can write
\begin{equation}\label{qBM2}
 S_{\rm qB}(x(t))=k \ln\dim {\cal H}_{M(t)} ,
\end{equation} 
where ${\cal H}_{M(t)}$ is the Hilbert space of the macroscopic state occupied by $x(t)$ at the time $t$.
This entropy also naturally grows with time if it is low initially,
essentially for the same reason as entropy (\ref{qBM}).
The entropy (\ref{qBM2}) is partially ontological and partially nomological. 

It is also instructive to compare the quantum entropies above with their classical cousins.
 
The von Neumann entropy (\ref{vNe}) is a quantum version of Gibbs entropy
\begin{equation}\label{Gibbs}
 S_{\rm G}(t) =-k \int dz\, \rho(z,t) \ln [\rho(z,t)\delta z] ,
\end{equation}
where $\rho(z,t)$ is the probability density in the classical phase space.
Note that probability {\em density} is not a dimensionless quantity,
so $\delta z$ is the volume of a very small 
``elementary'' cell in phase space, which ensures that $\rho(z,t) \delta z$ is dimensionless
and smaller than 1.   
Even though the entropy is classical,
the value of $\delta z$ is sometimes fixed to be equal to $h^{3N}$, where $h$ is the Planck constant.
But the actual value of $\delta z$ is not really important because different choices affect the entropy by an additive constant,
which does not affect the entropy {\em change} in physical processes, while only the change of entropy is of direct physical
relevance.  
In classical physics, $\rho(z,t)$ can be interpreted as a Bayesian probability, 
which is a tool to make probabilistic predictions 
about the behavior of particles with trajectory $z(t)$. Hence $\rho(z,t)$ encodes a probabilistic law, 
so Gibbs entropy defined by $\rho(z,t)$ is nomological. In open systems it tends to increase with time,
similarly to its quantum cousin,
because, if initially 
the open system is not much correlated with the rest of the Universe, the interaction with the 
rest of the Universe usually creates more correlation, which increases the Gibbs entropy.
In closed systems it doesn't change with time.
Nevertheless, even in closed systems Gibbs entropy can be used to describe the increase of entropy with time,
by replacing the fine grained entropy (\ref{Gibbs}) with its coarse grained version. 
In the coarse grained version, one replaces the continuous integral $\int dz$ 
with a discrete sum $\sum_z$ over a discrete set of finite cells, where each cell is labeled by its average $z$. 
   
It has been objected \cite{goldstein-BoltzGibbs-CQ} that Gibbs entropy is subjective, because the probability $\rho(z,t)$
is subjective. This objection may be correct in the Bayesian interpretation of probability, but not in the 
frequentist interpretation of probability. More interestingly, even if we keep the Bayesian interpretation of probability
and accept that Gibbs entropy is subjective, its quantum cousin von Neumann entropy is {\em not} subjective.
This is because the quantum density operator $\hat{\rho}(t)$ is obtained through a
mathematically well defined (and hence objective) partial trace of the objective pure state of the closed system.
In Bohmian mechanics, von Neumann entropy in this sense is not any less objective than the state $|\psi(t)\rangle$.
But it is an objective nomological entity, not an objective ontological entity.

Classical Boltzmann entropy, or simply Boltzmann entropy, is defined as
\begin{equation}\label{boltz}
 S_{\rm B}(\Gamma)=k \ln W(\Gamma) ,
\end{equation} 
where 
\begin{equation}\label{boltzW}
W(\Gamma)=\frac{1}{\delta z} \int_{\Gamma} dz
\end{equation}
is the dimensionless phase volume of a region $\Gamma$ in phase space. 
The small phase volume $\delta z$ of the ``elementary" cell has a role similar
to that in (\ref{Gibbs}). 
Similarly to the quantum Boltzmann entropy, 
it is particularly interesting when the full phase space is partitioned into phase-space cells 
analogously to (\ref{oplus})
\begin{equation}\label{cup}
 \Gamma = \bigcup_M \Gamma_M ,
\end{equation} 
where each cell $\Gamma_M$ corresponds to one macroscopic state that can be distinguished from other 
macroscopic states. Then one can define Boltzmann entropy as a function of the actual particle positions 
$z(t)$ in the phase space
\begin{equation}
  S_{\rm B}(z(t))=k \ln W(\Gamma_{M(t)}) ,
\end{equation}
where $\Gamma_{M(t)}$ is the cell occupied by $z(t)$ at the time $t$. If the Boltzmann entropy is low initially,
it naturally grows with time \cite{penrose,goldstein-BoltzGibbs-C,goldstein-BoltzGibbs-CQ},
because the prior probability that the particle is in the cell $\Gamma_M$ is proportional to $W(\Gamma_M)$, 
so growth of entropy is just a natural transition to a more probable macroscopic state.

It has been argued \cite{goldstein-BoltzGibbs-CQ,goldstein-BoltzGibbs-C} that Boltzmann entropy makes much more sense 
than Gibbs entropy. 
First, because Boltzmann entropy depends on the objective trajectory $z(t)$, rather than on the subjective (Bayesian)
probability density $\rho(z,t)$. Second, because Boltzmann entropy naturally grows with time. 
However, we have already explained that $\rho(z,t)$ can be interpreted objectively as frequentist probability, 
and that Gibbs entropy can naturally grow, either by coarse graining or by considering an open subsystem.
Moreover, Boltzmann entropy also involves a kind of subjectivity encoded in the choice 
of the partition (\ref{cup}). More fundamentally, similarly to the quantum Boltzmann entropy, the 
classical Boltzmann entropy can be made objective by introducing the Hamiltonian $H_O$ of the observer.
However, when the Boltzmann entropy depends on $H_O$, then it is not purely ontological, 
but partially nomological.

Note also that by ``Boltzmann entropy" people sometimes mean \cite{jaynes}
\begin{equation}\label{boltz_jaynes}
 S'_{\rm B} = -kN\int dz_1\, \rho_1(z_1,t) \ln [\rho_1(z_1,t)\delta z_1] ,
\end{equation}
where
\begin{equation}
\rho(z_1,t)=\int dz_2 \cdots \int dz_N\, \rho(z_1,z_2,\ldots,z_N,t)
\end{equation}
is the 1-particle probability density. The entropy (\ref{boltz_jaynes}) 
is really Gibbs-like, but can be derived 
from Boltzmann entropy (\ref{boltz}) as an approximation \cite{bain},
up to an unimportant additive constant depending on $\delta z_1$.

Last but not least, we point out that the advantage of Boltzmann entropy over Gibbs entropy 
is to a large extent lost when they are replaced with their quantum cousins.  
This is because,
as we explained above, von Neumann entropy is ``more objective'' than Gibbs entropy, 
because $\hat{\rho}(t)$ is objective in a sense in which classical $\rho(z,t)$ isn't.

Finally, let us note that there is a notion of Gibbs-like entropy that deals directly with statistics 
of Bohmian particle positions, which has been applied to explain 
why the Bohmian particle positions have the probability density equal to $\rho(x,t)$ \cite{valentini}. 
Unlike other entropies discussed above, this type of entropy does not play any direct role 
in standard quantum statistical mechanics. In particular, this type of entropy does not give 
rise to thermal entropy. Ultimately, this is because thermal entropy is related to the Hamiltonian of the 
system, while Bohmian mechanics is not Hamiltonian mechanics. 
In this paper, this type of entropy will not be used.

To conclude, there are many kinds of entropy and each has its advantages and disadvantages. 
There is no such thing as the ``right'' definition of entropy.
Their roles are complementary, different kinds of entropy are relevant in different contexts.
We shall also further discuss it in Sec.~\ref{SECmix}.

\subsection{Explanation of thermodynamics}

In this subsection, most of the technical results are already known from existing literature 
and do not depend on Bohmian interpretation. Hence we shall not present detailed 
derivations, but only sketch the main ideas, with emphasis on conceptual ideas which are essential 
from the Bohmian point of view.

In realistic systems described by statistical mechanics, the mixed state usually originates from partial tracing
of the state of a larger system. Thus the thermal mixed state (\ref{Srho}) is really an example of the state 
$\hat{\rho}_A$ obtained by partial tracing over the rest of the larger system with the state $\hat{\rho}$. In other words,
(\ref{Srho}) can be written more correctly as 
\begin{equation}\label{Srho3}
 \hat{\rho}_A(t)={\rm Tr}_B\hat{\rho}(t) \simeq \frac{e^{-\beta \hat{H}_A}}{ {\rm Tr}\, e^{-\beta \hat{H}_A} } ,
\end{equation}
where $\hat{H}_A$ is the Hamiltonian of the $A$-subsystem. 
(Alternatively, (\ref{Srho3}) can also be obtained by maximizing the von Neumann entropy under the constraints 
that the density matrix is normalized and that the ensemble average of Hamiltonian has a certain prescribed value \cite{sakurai},
but we shall not pursue this approach.) 

A more detailed explanation of the approximate equality in (\ref{Srho3}) is as follows.
One starts from a big system, called thermal bath, in a microcanonical ensemble, meaning that its energy 
is uniformly distributed within a narrow range $[\bar{E},\bar{E}+\Delta E]$, where $\Delta E$ is small.
Thus the bath can be modeled by a pure state
\begin{equation}\label{psimc}
|\psi(t)\rangle \propto \sum_{{\rm microcan} \; m} e^{-iE_mt/\hbar} |m\rangle , 
\end{equation}
where ``${\rm microcan} \; m$'' denotes that the sum is taken only over states the energy of which satisfies
$E_m\in[\bar{E},\bar{E}+\Delta E]$.
The big system has $N\gg 1$ particles
and its density operator is $\hat{\rho}(t)=|\psi(t)\rangle\langle\psi(t)|$.
Now consider an arbitrary subsystem of $N_A$ particles, such that 
\begin{equation}
 N\gg N_A\gg 1.
\end{equation}
We assume that the $A$-subsystem weakly interacts with the rest of the bath (the $B$-subsystem),
such that the full Hamiltonian has the form $\hat{H}=\hat{H}_A+\hat{H}_B+\hat{H}_{AB}$ 
and the $A$-subsystem is entangled with the $B$-subsystem, but in a computation of the full energy 
of the system one can use the approximation $\hat{H} \simeq \hat{H}_A+\hat{H}_B$.
It turns out \cite{goldstein_canonical,popescu}
that almost any $A$-subsystem that satisfies the assumptions above has the mixed density matrix 
that can be approximated by (\ref{Srho3}). The $\beta$ in (\ref{Srho3}) is given by the formula
\cite{goldstein_canonical}
\begin{equation}
 \beta=\frac{1}{k} \frac{dS_{\rm qB}(\bar{E})}{d\bar{E}} ,
\end{equation}
where $S_{\rm qB}(\bar{E})$ is the quantum Boltzmann entropy of the full system in the microcanonical ensemble
\begin{equation}
 S_{\rm qB}(\bar{E})=k \ln\dim {\cal H}_{{\rm microcan}}(\bar{E}) ,
\end{equation}
and $\dim {\cal H}_{{\rm microcan}}(\bar{E})$ is the dimension of the microcanonical Hilbert space,
spanned by all states the energy of which is in the interval $[\bar{E},\bar{E}+\Delta E]$. 
For more details we refer the reader to \cite{goldstein_canonical}.

The derivation of (\ref{Srho3}) from (\ref{psimc}) explains how canonical ensemble of a subsystem arises 
from the microcanonical ensemble of the full closed system. After that, one can proceed as in standard 
statistical mechanics textbooks, which we now briefly sketch. It will be understood that everything is considered 
in the $A$-subsystem, so the label $A$ will be suppressed.  
One starts from introducing the partition function 
\begin{equation}
 Z(V,T)={\rm Tr}\, e^{-\beta \hat{H}} = \sum_n e^{-\beta E_n} ,
\end{equation}
where $\beta=1/(kT)$ as usual. The dependence on the spatial 3-dimensional volume $V$ arises because 
$\hat{H}= \hat{H}(V)$ describes a system confined within $V$, so that wave functions which are eigenstates of
$\hat{H}(V)$ all vanish outside of $V$. The density operator (\ref{Srho3}) 
is diagonal in the $n$-basis, and its eigenvalues
\begin{equation}
 p_n=\frac{e^{-\beta E_n}}{Z}
\end{equation}
satisfy $\sum_n p_n =1$. 
Thus the average energy (\ref{Eaverage}) can be written as
\begin{equation}\label{Etherm}
 E=\sum_n E_n p_n =
\frac{\sum_n E_n e^{-\beta E_n}}{Z} = -\frac{\partial \ln Z}{\partial\beta}=kT^2\frac{\partial \ln Z}{\partial T}.
\end{equation}
Similarly, the entropy (\ref{vNe}) can be written as 
\begin{equation}\label{Stherm}
 S \equiv S_{\rm vN} =-k\sum_n p_n \ln p_n = \frac{\partial (kT \ln Z)}{\partial T} .
\end{equation}
Defining also the pressure as 
\begin{equation}\label{Ptherm}
 P=\frac{\partial (kT \ln Z)}{\partial V} ,
\end{equation}
by standard thermodynamic methods \cite{huang,reif} one finds\footnote{A quick reminder for those 
whose knowledge of thermodynamics is rusty. From (\ref{Stherm}) and (\ref{Etherm})
one finds $S=E/T+k\ln Z=(E-F)/T$, where $F(V,T)\equiv -kT\ln Z(V,T)$ is Helmholtz free energy.
Hence $E=F+TS$, so $dE=(\partial F/\partial V)dV+(\partial F/\partial T)dT + TdS + SdT$.
Using (\ref{Ptherm}) and (\ref{Etherm}), this can be written as 
$dE=-PdV+(F-E)dT/T + TdS + SdT = TdS -PdV +(F-E+TS)dT/T$. The last bracket is zero, 
which proves (\ref{1stlaw}).}
that these quantities obey 
\begin{equation}\label{1stlaw}
 dE=TdS-PdV ,
\end{equation}
which is nothing but the 1st law of thermodynamics. 

So far, in this subsection we said nothing new and used only standard quantum statistical mechanics. 
Now we interpret (\ref{1stlaw}) from the Bohmian point of view. For that purpose, 
we stress that all quantities in (\ref{1stlaw}) are purely nomological, they say nothing about 
Bohmian ontology. The crucial Bohmian insight arises from the general result that Bohmian particle positions in space 
are distributed in the same way as in standard QM. Hence, the volume $V$ inside which the eigenfunctions 
of $\hat{H}$ do not vanish is the same as the volume $V_{\rm Bohm}$ filled with actual Bohmian particles.
In other words, Bohmian mechanics predicts that
\begin{equation}
   V_{\rm Bohm} = V ,
\end{equation}
so (\ref{1stlaw}) implies
\begin{equation}\label{1stlawB}
 dE=TdS-PdV_{\rm Bohm} ,
\end{equation}
which is the Bohmian version of the 1st law of thermodynamics. An even more illuminating way to write this is
\begin{equation}\label{1stlawB2}
 \frac{dV_{\rm Bohm}}{dt}=\frac{1}{P} \left( T \frac{dS}{dt} - \frac{dE}{dt} \right) ,
\end{equation}
which has the form of an equation of motion for the ontological thing on the left-hand side,
guided by the nomological thing (the law) on the right-hand side. 
This is, in essence, how Bohmian mechanics explains thermodynamics.

We note that (\ref{1stlawB2}) describes a change of volume of a macroscopic object (typically a gas),
so it is not necessary to introduce an additional macroscopic measuring apparatus to 
relate (\ref{1stlawB2}) with directly observable things. The macroscopic volume $V_{\rm Bohm}$ 
is itself {\em directly} observable, it is a perceptible in the language of \cite{nikIBM}.
It should be distinguished from the microscopic Bohmian law of motion (\ref{traj}) or (\ref{trajAt}),
which is not observable directly.

From the Bohmian point of view, the most important and perhaps unexpected conceptual message 
is that most thermodynamic quantities are {\em not} ontological. Even though Bohmian particles have actual positions,
actual velocities, and hence actual kinetic energies, the energy $E$ appearing in the 1st law of thermodynamic 
does not depend on those actual positions and velocities, so this energy is not ontological.
Likewise, the temperature $T$ also does not depend on those actual positions and velocities,
so temperature is also not ontological. Similarly, the entropy $S$ also does not depend on those actual 
positions and velocities, so it is also not ontological. Instead, all these quantities are nomological, they are entities
that describe the law of motion for the ontological particles.
The only ontological thermodynamic quantity is the volume
$V_{\rm Bohm}$ (numerically equal to the nomological volume $V$), corresponding to the space filled with 
ontological particles. We find this message very deep and important, because otherwise someone with a Bohmian way of thinking 
can easily fall into a trap of conceptual confusion by trying to think of energy, temperature, entropy and pressure 
as ontological quantities somehow to be defined by the actual Bohmian positions and velocities.  
The thermodynamic quantities $E$, $T$, $S$, $P$, and even $V$, are the quantities specified  
by the mixed quantum state $\hat{\rho}$, so they are nomological, 
in the same sense in which the wave function is nomological, and not ontological.
 
\subsection{Proper mixture {\it vs} improper mixture, and the meaning of entropy}
\label{SECmix}

In quantum foundations one often distinguishes a proper mixture 
from an improper mixture \cite{despagnat}.
In a proper mixture, the subsystem of interest is actually in a pure state, but one does not know 
what this pure state is, so one describes the subsystem statistically in terms of a mixed
density operator. In an improper mixture, the mixed density operator of the subsystem originates 
from partial tracing over the rest of the Universe, so there is no single pure state 
that could be associated with the subsystem. However, in practical applications 
of standard quantum statistical mechanics, one rarely cares about the difference, because the 
average values such as (\ref{Eaverage}) do not depend on it. What can Bohmian mechanics 
say about the difference between proper and improper mixtures?

In our formulation, the density operator $\hat{\rho}_A(t)$ is always obtained through partial tracing over the rest 
of the closed system, so $\hat{\rho}_A(t)$ is an improper mixture. However, the second line of (\ref{sumtj})
tells us that the current operator 
of the $A$-subsystem can be expressed in terms of $\hat{\rho}_A(t)$, 
in a way which does not see a difference between proper and improper mixtures. The consequence is that the 
effective Bohmian trajectories (\ref{trajAt}) do not depend on whether the mixture is proper or improper.
On the other hand, the more fundamental trajectories (\ref{traj}) take the whole 
closed system into account, so the mixed density operator of the $A$-subsystem, proper or improper, 
does not take any role in a computation of Bohmian trajectories.    
In any case, the measurable predictions of Bohmian mechanics do not depend on whether 
one uses (\ref{trajAt}) or (\ref{traj}). Hence, for all practical purposes, 
the difference between proper and improper 
mixtures in statistical physics of {\em large} systems is irrelevant. 

Let us illustrate it on an example. Consider a cat that can be in two macroscopically different states, cold cat and warm cat.
Clearly, the thermal entropy $S_{\rm warm}$ of the warm cat is larger than the thermal entropy $S_{\rm cold}$ of the cold cat. 
But what if the cat is in the mixture of cold and warm? 
And what does it even {\em mean} that the cat ``is in the mixture''?

For the sake of argument, let us first assume that the cat can be sufficiently isolated from the environment, so that
the cold cat can be modeled by a closed system in the pure state $|\rm cold \; cat\rangle$, and similarly for the warm cat 
in the pure state $|\rm warm \; cat\rangle$. This, of course, cannot be realized in practice, but it should be possible in principle.
So in principle, it should also be possible to have a coherent superposition 
\begin{equation}
 \frac{1}{\sqrt{2}} [|\rm cold \; cat\rangle + |\rm warm \; cat\rangle ] .
\end{equation}
The two wave functions $\langle x|\rm cold \; cat\rangle$ and $\langle x|\rm warm \; cat\rangle$ have a negligible 
overlap in the $x$-space, so they correspond to two separated channels, which implies that the Bohmian positions $x(t)$ 
will be either in the cold channel or the warm channel. In other words, the cat is either cold or warm, and the 
thermal entropy of the cat is either $S_{\rm cold}$ or $S_{\rm warm}$. 

But since the cat is isolated from the environment, an external observer does not know whether the cat is cold or warm.
Hence the observer can describe his knowledge by the {\em proper} mixture, represented by the mixed density operator
\begin{equation}\label{mixcat}
  \frac{1}{2} [|{\rm cold \; cat}\rangle\langle{\rm cold \; cat}| + |{\rm warm \; cat}\rangle\langle{\rm warm \; cat}| ] .
\end{equation}
What is the entropy of this state? The answer depends on what one {\em means} by ``entropy''.
The thermal entropy is still either $S_{\rm cold}$ or $S_{\rm warm}$. The von Neumann entropy is $k\ln 2$.

Now suppose that a measuring apparatus measures the cat, so that the state of the cat gets entangled with the state 
of the measuring apparatus
\begin{equation}
 \frac{1}{\sqrt{2}} [|{\rm cold \; cat}\rangle |B_{\rm cold}\rangle  + |{\rm warm \; cat}\rangle |B_{\rm warm}\rangle] ,
\end{equation} 
where $|B_{\rm cold}\rangle$ and $|B_{\rm warm}\rangle$ are the possible states of the measuring apparatus. 
But suppose that the observer does {\em not know} the result of measurement. This means that the observer 
will describe the cat by the {\em improper} mixture, represented by the {\em same} density operator (\ref{mixcat}).
The observer will associate a proper mixture with the full cat+apparatus system, but the cat alone 
is described by an improper mixture. 
The thermal entropy is still either $S_{\rm cold}$ or $S_{\rm warm}$, and the von Neumann entropy is still $k\ln 2$.

Now consider a more realistic situation, in which the cat has never been isolated from the environment. 
Instead, the cat is an $A$-subsystem of a big thermal bath. The bath has a temperature, 
the value of which is either $T_{\rm cold}$ or $T_{\rm warm}$.
Hence the state of the cat is either $\hat{\rho}_{\rm cold}$ or $\hat{\rho}_{\rm warm}$,
each of which is an {\em improper} mixture represented by a density operator of the form (\ref{Srho3}),
with $\beta=\beta_{\rm cold}$ or $\beta=\beta_{\rm warm}$. But if an observer does not know 
the temperature, he will describe his knowledge about the cat with a mixture 
\begin{equation}
 \frac{1}{2} [\hat{\rho}_{\rm cold} + \hat{\rho}_{\rm warm}] .
\end{equation}
Is this mixture proper or improper? The question is tricky. It is ``proper'', 
in the sense that the cat is either in the state $\hat{\rho}_{\rm cold}$ or  $\hat{\rho}_{\rm warm}$.
But it is also ``improper'', in the sense that $\hat{\rho}_{\rm cold}$ and $\hat{\rho}_{\rm warm}$
themselves are improper mixtures. And what about entropy? The thermal entropy is either 
$S_{\rm cold}$ or $S_{\rm warm}$. The von Neumann entropy, on the other hand, is approximately the average
of $S_{\rm cold}$ and $S_{\rm warm}$ \cite{goldstein-BoltzGibbs-CQ}. 

It has been argued in \cite{goldstein-BoltzGibbs-CQ} that the ``true'' entropy is simply $S_{\rm cold}$ or $S_{\rm warm}$,
from which the authors concluded that von Neumann entropy, and its classical cousin Gibbs entropy, are ``wrong''.
They used it as one of the arguments for the general idea that only Boltzmann entropy (classical or quantum)
is ``right''. However, this argument assumes that there {\em is} such thing as ``the right entropy''.
In our opinion, there is no such thing. After all, at the ontological level, the Bohmian particles 
only have trajectories $x(t)$. They don't have entropy {\em at all} at the ontological level.
As we discussed in Sec.~\ref{SECentropy}, all kinds of entropy are at least partially nomological.
In particular, Boltzmann entropy is nomological because it depends on the Hamiltonian of the 
observer $H_O$, which determines a ``natural'' partition of the phase (or Hilbert) space into 
macroscopically distinguishable subspaces. 
More importantly, different kinds of entropy are just different thinking tools that, in one way or another,
help us to effectively describe complex phenomena in simple terms. 
It makes no sense to ask what is ``the right entropy'' before specifying what one wants to do with it.
The world out there does not {\em have} entropy; it is us, the scientists, who ascribe {\em an} entropy
(not {\em the} entropy) to it, depending on the level of description that we use in a given context.   

\section{Conclusion}
\label{SECconc}

To a large extent, quantum statistical mechanics reduces to a study of certain mixed density operators $\hat{\rho}$,
which usually arise through a partial trace over the environment. Thus, to a large extent, 
quantum statistical mechanics can be thought of as a study of certain open subsystems. On the other hand, Bohmian mechanics
in its usual formulation insists that, in principle, the whole closed system must always be taken into account, 
thereby somewhat contradicting the general spirit of quantum statistical mechanics. 

In this paper we have developed a new 
approach to Bohmian mechanics, which can treat an arbitrary subsystem (with a fixed number of particles)
without any reference to the rest of the system, thus making it more convenient for applications in quantum statistical mechanics.
The truncated Bohmian particle trajectories in the subsystem obtained in this way differ from Bohmian trajectories
in the subsystem obtained by considering the full system. Nevertheless, the trajectories are not directly measurable and the 
probabilistic predictions obtained by the two approaches are the same, and equivalent to probabilistic 
predictions obtained by standard quantum theory. The purpose of Bohmian trajectories in our approach
is not to postulate the absolute ``true reality'', but to construct an intuitive picture useful for conceptual thinking
about quantum phenomena. According to this picture, the world is made of particles which always have definite positions,
irrespective of whether they are measured or not, in a way which is compatible with all probabilistic 
predictions of non-relativistic quantum mechanics. Moreover, the particle trajectories are deterministic, while all probabilities 
emerge from a lack of knowledge of actual initial conditions, very much like in classical statistical mechanics.
In this way, with the Bohmian interpretation, quantum statistical mechanics is conceptually very similar
to classical statistical mechanics. The differences between classical and quantum statistical mechanics 
can be reduced to the fact that, for closed systems, classical particle trajectories are determined from the classical Hamiltonian $H$,
while Bohmian particle trajectories are determined from the density operator $\hat{\rho}$.
  
The novel results in this paper are both technical and conceptual.
On the technical side, we have developed the formalism of probability currents in the multi-particle space
of open systems, for both classical and quantum mechanics. The currents, obeying continuity equations, 
define natural velocities of particles, compatible with probabilities in the considered open system. 
On the conceptual side, we have developed a version of Bohmian interpretation in which 
particle trajectories are interpreted merely as an intuitive thinking tool, without any pretensions 
to claims about reality of these trajectories. We have also discussed how Bohmian trajectories help 
in conceptual thinking about various notions of entropy, about proper and improper mixtures, and about thermodynamics.
   
To our knowledge, this work is the first systematic study of quantum statistical mechanics from a Bohmian point of view.
We certainly did not answer all possible questions about the role of Bohmian interpretation in statistical mechanics, 
so we hope that this work will stimulate further research.

\section*{Acknowledgments}
The author is grateful to X. Oriols for very detailed discussions on the role of 
Bohmian mechanics in quantum statistical mechanics, and for reading the manuscript.

\end{document}